\documentclass[aps,apl,amsmath,amssymb,twocolumn]{revtex4}
\usepackage{graphics}
\usepackage{graphicx}
\usepackage{subfigure}
\usepackage{epsfig}
\usepackage{amssymb}
\usepackage{amsthm}
\usepackage{amsmath}
\usepackage{epstopdf}
\usepackage{setspace}
\usepackage{subfigure}
\usepackage{longtable}
\usepackage{color}

\begin{document}

\title{Optimization of optical absorption in thin layers of amorphous silicon enhanced by silver nanospheres}

\author{Mikhail Omelyanovich$^{1,2}$, Younes Ra'di$^1$ and Constantin Simovski$^{1,2}$}
\address{$^1$Department of Radio Science and Engineering, Aalto University, P.O. Box 13000, FI-00076, Finland}
\address{$^2$Laboratory of Metamaterials, University for Information Technology, Mechanics and Optics (ITMO), St. Petersburg 197101, Russia}

\begin{abstract}
We study a highly controllable perfect plasmonic absorber -- a thin metamaterial layer which possess balanced electric and magnetic responses
in some frequency range. We show that this regime is compatible with both metal-backed variant of the structure or its semitransparent variant. This regime 
can be implemented in a prospective thin-film photovoltaic cell with negligible parasitic losses.
\end{abstract}


\maketitle

Thin-film photovoltaic (PV) cells enhanced by light-trapping structures (LTSs), capable to prevent both reflection from the cell and parasitic transmission through its PV layer \cite{TFSC}
have not yet been adapted by the industry in spite of rather long history of corresponding research started by work \cite{APL}.
Perhaps, this is so due to the lack of efficiency of suggested LTSs and/or impractical design solutions in the cases when these LTSs are efficient enough.
Most popular LTSs are those based on plasmonic nanostructures which allow the incident light to concentrate in strongly subwavelength regions inside the PV layers \cite{Polman}.
These structures were, however, heavily criticized for parasitic losses in the metal nanoelements (see e.g. in \cite{Akimov}). Really, high losses in the metal constituents make the light-trapping functionality meaningless. 
Without an LTS some incident power is absorbed in the bottom electrode, in presence of a plasmonic LTS it can be absorbed inside the latter one, which is not better. Besides of the direct losses of the incident light energy this implies a decrease of the PV conversion efficiency due to the heating of the PV layer \cite{TFSC}.

\begin{figure}[ht]
\centering
\includegraphics[width=7.5cm]{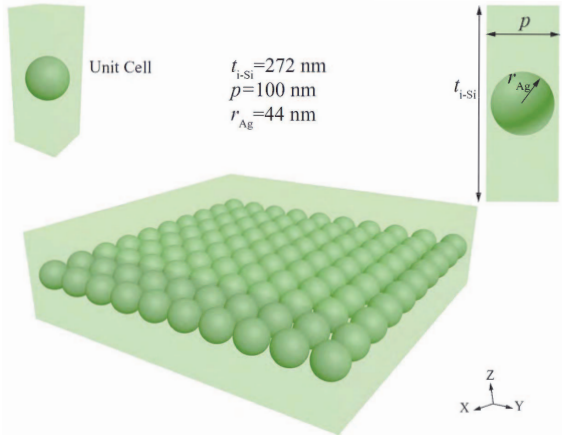}
\caption{Amorphous silicon layer with symmetrically located grid of densely packed plasmonic nanospheres and the unit cell of the structure.} \label{fig1}
\end{figure}
 
Some regular arrays of metal nanoelements called plasmonic nanoantennas are capable to concentrate the light aside them. At the corresponding frequencies the parasitic losses become negligible 
\cite{Ferry,Brongersma,Simovski,Sinev}. However, such structures seem to be expensive in fabrication, and their practical prospectives in thin-film photovoltaics are disputable.  The same can be said on 
regular plasmonic arrays converting an incident plane wave into waveguide modes of the PV layer (i.e. mimicking an optical facet which cannot be directly used for a thin film due to the so-called Yablonovitch limit) \cite{Spinelli}.

\begin{figure}[ht]
\centering
\includegraphics[width=5cm]{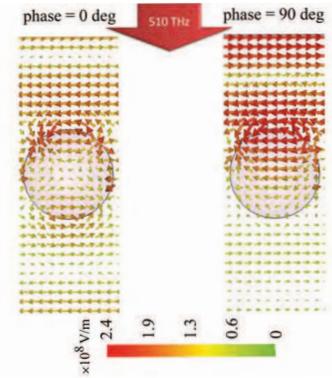}
\caption{Electric field distribution at the operation frequency 510 THz for two time moments shifted by one quarter of the period.} \label{fig2}
\end{figure}

The structure recently studied in paper \cite{1} is an optically thin metamaterial layer (such layers are often called metasurfaces) formed by simple plasmonic nanospheres. In Fig. \ref{fig1} we show its analogue where the n-doped a-Si is replaced by intrinsic a-Si. The structure represents a kind of so-called perfect plasmonic absorber (PPA) (see e.g. in overview \cite{4}). However, this PPA is rather unusual -- it implements the so-called Huygens' metasurface \cite{4,5,6,7}, i.e., possessing balanced electric and magnetic dipole responses of a unit cell. In this the regime the reflection is damped, and the metasurface re-radiates only along the incidence direction. This radiation is, however, not transmitted though the metasurface being fully absorbed.

\begin{figure}[ht]
\centering
\includegraphics[width=7.5cm]{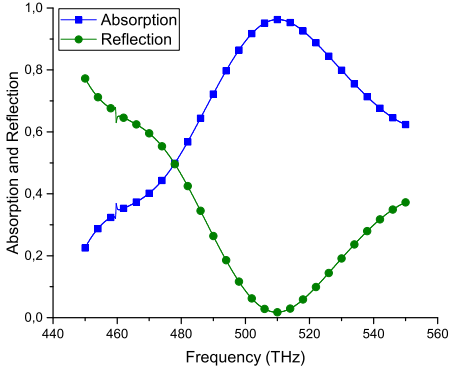}
\caption{Absorption and reflection versus frequency when the host material is intrinsic a-Si.} \label{fig3}
\end{figure}

The magnetic resonance revealed in \cite{1} is different from the magneto-dipole Mie resonance observed in refractive particles \cite{7}, from the magnetic response of an asymmetric Huygens' metasurface observed along with the so-called substrate-induced biansiotropy \cite{5}, and from the magnetic resonance of an optical fishnet \cite{8}. As well as in the arrays of nanoantennas \cite{Sinev} this resonance corresponds to a collective mode of the array. It is illustrated by the distribution of local electric field vectors $\bf E$ (polarization current density is proportional to $\bf E$ with the coefficient $i\omega(\varepsilon-1)$) in the bulk of the unit cell depicted  in Fig. \ref{fig2}. Two color maps in Fig. \ref{fig2} are simulated for two time moments shifted by a quarter of the period from one another. Figure \ref{fig2} allows us to find the magnetic mode region. Though $\bf E$ concentrated in the top half of the layer is polarized nearly horizontally, it is not the field of a plane wave transmitted into a-Si. First, vectors $\bf E$ change their direction to the opposite one so that the planes of maximal intensity are distanced by nearly 100 nm. This is noticeably smaller than the half-wavelength (distance needed for the $\pi$ phase shift of the plane wave) in a-Si at frequency 500 THz.  Second, the concentration of polarization currents in the top part of the layer disappears when Ag nanospheres are removed.

Identifying the domain of the magnetic mode as the top half of the layer one can calculate the magnetic moment of the unit cell using its general definition as it was done \cite{1}. The electric dipole moment is calculated by a simple integration of polarization currents. Notice, that the electric response at 500 THz is also a collective mode: the dipole plasmon resonance of an inidivdual nanosphere and its high-order Mie resonances are located beyond the frequency region of our interest. At 500-502 THz the electric and magnetic moments are balanced and the structure represents the Huygens' metasurface. 

\begin{figure}[ht]
\centering
\subfigure[]{\includegraphics[width=6cm]{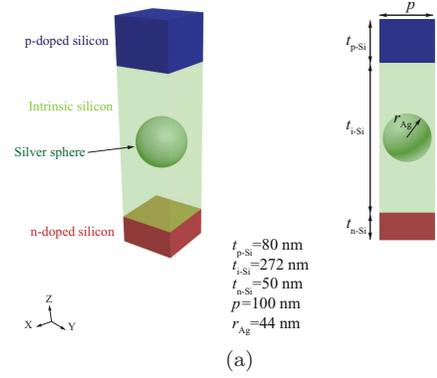}}
\subfigure[]{\includegraphics[width=6cm]{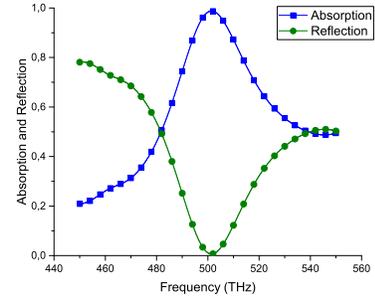}}
\subfigure[]{\includegraphics[width=6cm]{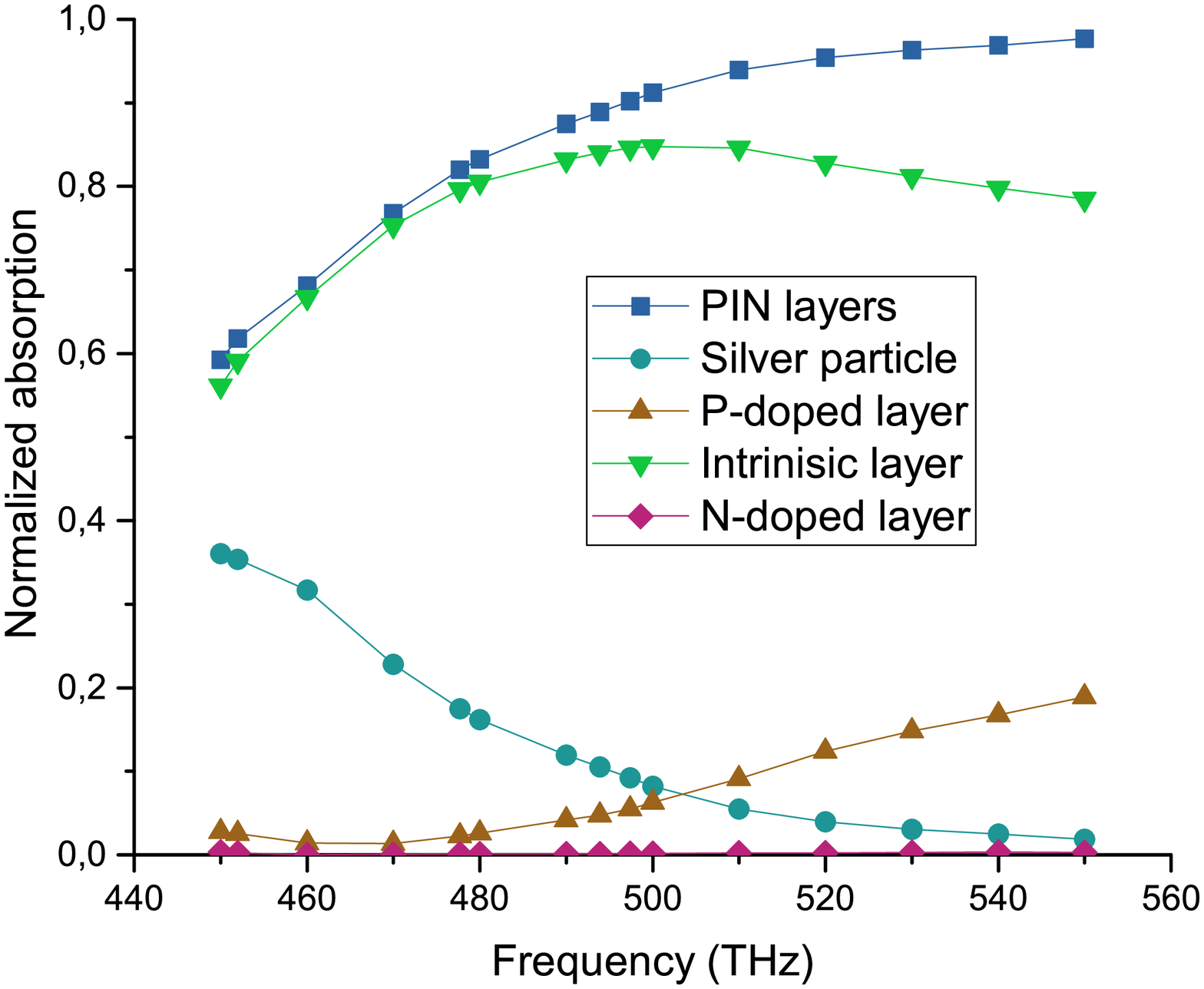}}
\caption{The p-i-n structure with Ag spheres in the intrinsic a-Si (a). Total power absorption and reflection versus frequency for normal incidence (b). Absorption in all constituents of the structure (c).} \label{fig4}
\end{figure}

In the present paper we reveal and explore two unusual features of this Huygens' regime (combined with nearly perfect absorption) which make it promising for thin-film photovoltaics. First, we notice that the electric field rather weakly penetrates into nanospheres and the absorption holds mainly in the host medium. Since the n-doped a-Si studied in \cite{1} has low PV spectral response in the visible range, we replaced it by intrinsic a-Si whose PV response is high. This change allows the same Huygens' regime as in \cite{1} with slightly shifted operation frequency (from 500 THz to 510 THz) and slightly changed geometric parameters, see Fig. \ref{fig1}. The nearly PPA regime is illustrated in Fig. \ref{fig2} where the power absorption $A$ and reflection $R$ coefficients are shown versus frequency for the normal incidence. At the resonance the absorption attains 97\%. The reliability of simulations is ensured by the application of two solvers as in our work \cite{1}. The data for Ag and a-Si were taken from \cite{Johnson} and \cite{Stut}, respectively.

\begin{figure}[ht]
\centering
\subfigure[]{\includegraphics[width=6cm]{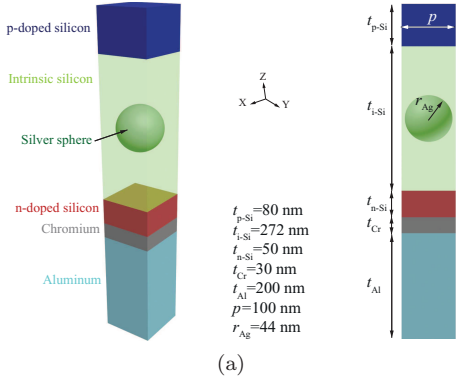}}
\subfigure[]{\includegraphics[width=6cm]{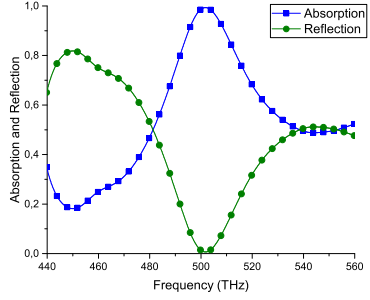}}
\subfigure[]{\includegraphics[width=6cm]{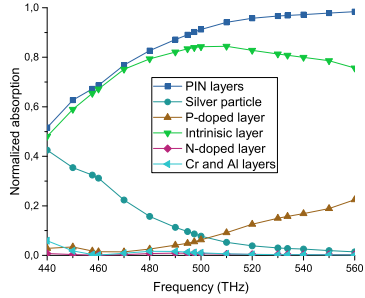}}
\caption{The metal-backed p-i-n structure with Ag spheres. Total absorption and reflection versus frequency (b). Absorption in all constituents of the structure (c).} \label{fig5}
\end{figure}

The second useful feature of the Huygens' regime revealed in \cite{1} is low value of the polarization current in the bottom part of the layer. Since the field almost does not penetrate there we may arbitrarily vary the material in this bottom part. On the next step we transit from the uniform host to a p-i-n structure which is mostly used in PV cells based on a-Si \cite{TFSC}. We have introduced a 30-nm thick layer of p-doped a-Si on top of the structure and a 25-nm thick heavily n-doped layer on the bottom. This structure is shown in Fig. \ref{fig4}(a). In spite of high optical losses in the bottom layer the useful regime survives. We may see in Fig. \ref{fig4}(b), that the resonant absorption becomes total (higher than 99.9\%) at 502 THz. In  Fig. \ref{fig3}(c) where the contributions of all constituents into overall absorption  \ref{fig4}(c) are shown we see that at the operation frequency the parasitic losses in Ag are as small as 4\%, whereas parasitic losses in the bottom layer of n-doped silicon are fully negligible. This implies the strong frequency selectivity of the useful absorption (the absorption in the p-layer also contributes into photocurrent). So, our structure can be used for a very efficient and ultimately thin PV cell. Since the optimal regime is frequency selective (bandwidth close to 10\%) this is not a solar cell, but
a PV cell operating in an artificial light. Such cells can be, for example, applied on top of biomedical chips which may be laminated by a PV film \cite{new}. Then the power for feeding the chip is harvested from a collimated narrow-band light.

\begin{figure}[ht]
\centering
\subfigure[]{\includegraphics[width=6cm]{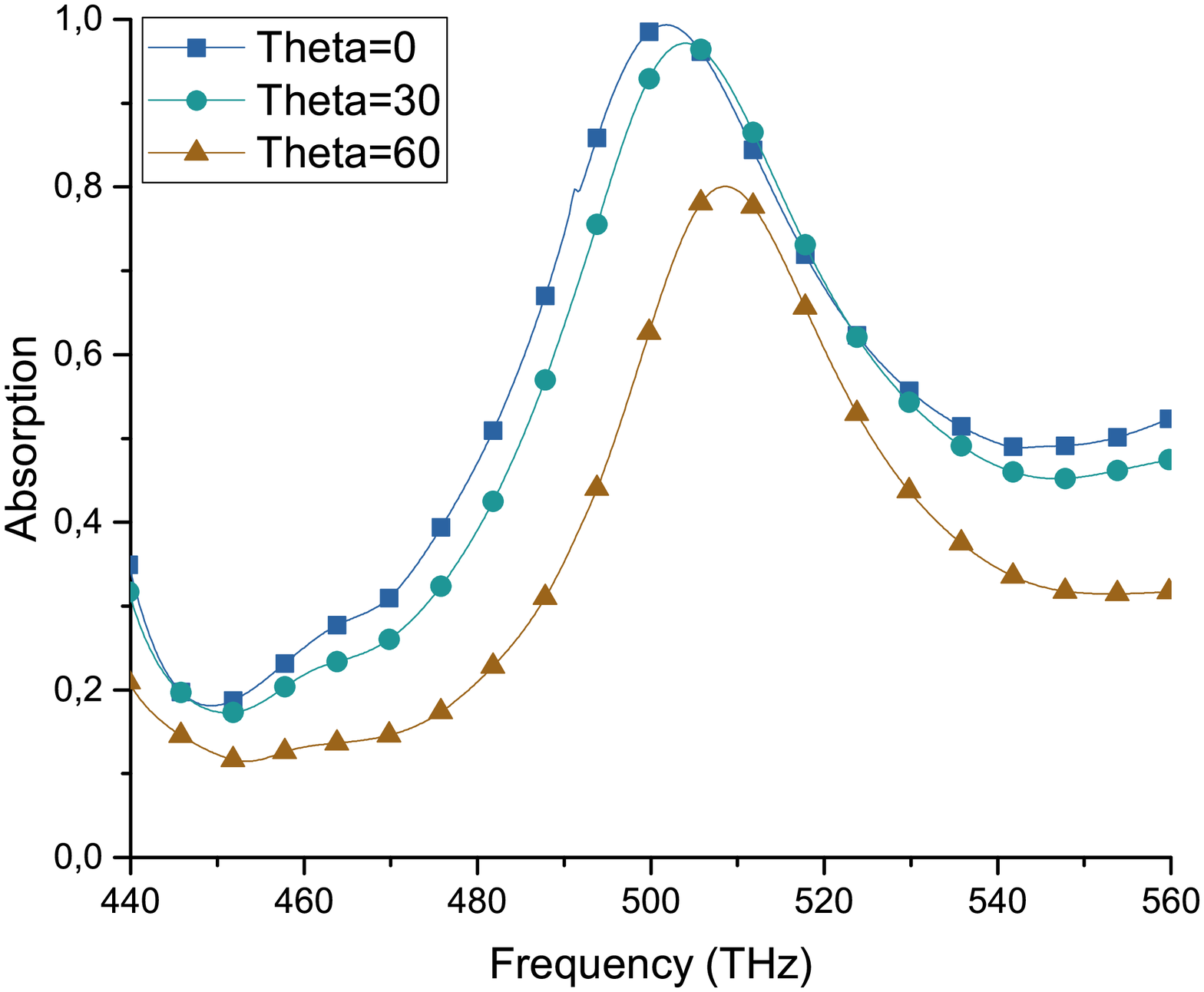}}
\subfigure[]{\includegraphics[width=6cm]{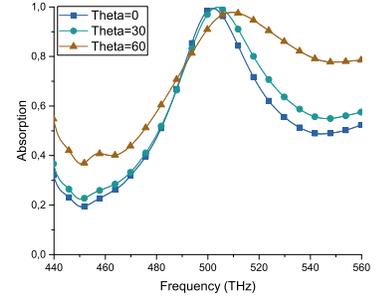}}
\caption{Absorption for the TE (a) and TM (b) cases of the oblique incidence of linearly polarized light on the structure from Fig. \ref{fig4}(a), incidence angles are $\theta = 0, 30, 60^{\circ}$.} \label{fig6}
\end{figure}

Though our metasurface is very thin and the fraction of the metal is nearly equal 0.75, it is weakly transparent: the transmittance $T=1-A-R$ does not exceed 3\% in the whole range under study (440-560 THz).
If the incident light is narrowband there is no need to keep the structure semi-transparent and the rear contact of our PV cell can be made of metal. In Fig. \ref{fig5}(a) it is a thin layer of chromium 
(needed in order to ensure the ohmic conductance of the transition semiconductor-metal) on the aluminum substrate. From plots in Fig. \ref{fig5}(b) simulated for this structure it is seen that the optimal regime 
for this structure shifts back to the range 500-502 THz. Fig. \ref{fig5}(c) shows that the absorption in the intrinsic layer of a-Si still dominates, and the absorption in Ag increases only slightly attaining 7\% at the operation frequencies. In all these simulations we keep in mind a top current-collecting electrode performed as a sparse wire mesh \cite{TFSC} which creates 5-10\% shadow losses but cannot influence to the absorption in a microscopic unit cell. The presence of such top electrode does not affect our simulations.

\begin{figure}[ht]
\centering
\subfigure[]{\includegraphics[width=6cm]{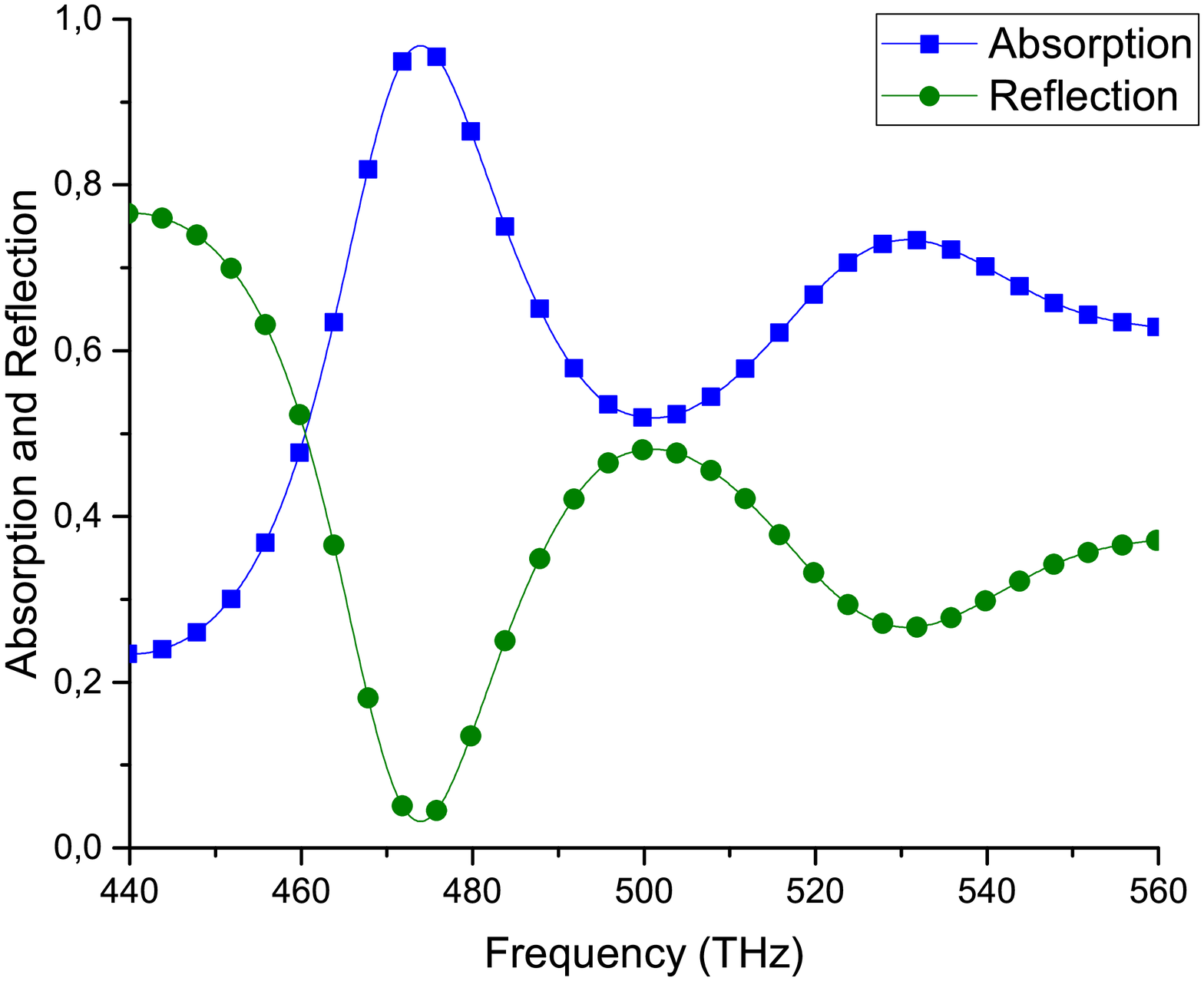}}
\subfigure[]{\includegraphics[width=6cm]{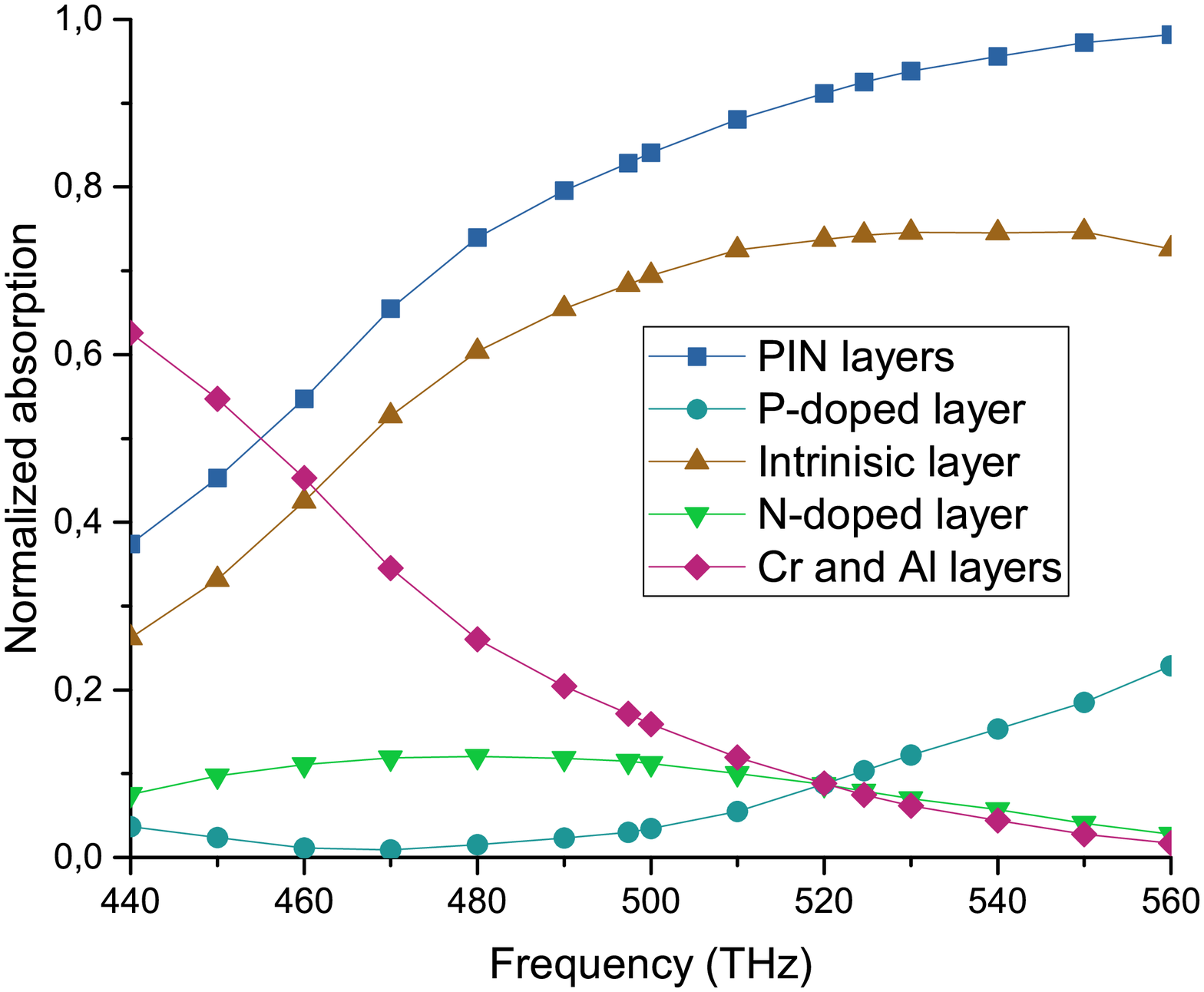}}
\caption{Absorption and reflection of structure without silver sphere inside intrinisic layer. Absorption in all constituents of the structure(c).} \label{fig7}
\end{figure}

Next, we have studied how the absorption in the semiconductor varies versus the angle of incidence. The simulations of $A(\omega)$ for TE and TM polarized light are illustrated by Fig. \ref{fig6}.
The optimal regime in the same frequency range keeps until 30$^{\circ}$ where the absorption is close to total and is weakly sensitive to the light polarization.

An important part of our study is the comparison of our light-trapping regime with the anti-reflecting regime which does not require Ag nanospheres. Our structure with removed nanospheres has very low reflection (4\%) at 473-475 THz. At these frequencies the transmission of the power through the PV layer is rather high as 30\%. Moreover, in the n-doped layer 10\% of the incident power is absorbed. It is clear that 40\% for the parasitic absorption (fully spent to the heating) is not acceptable for the selected application. So, the light-trapping regime is much better than the anti-reflecting regime, and the use of Ag nanospheres is justified. This conclusion does not change if one adds an anti-reflecting coating (e.g. a blooming layer of silica) since it does not prevent the plane-wave transmission through the PV layer. 

We have found that the same absorption regime keeps if the thickness of the intrinsic silicon is reduced from 272 nm to 184 nm. Since the field weakly penetrates into the bottom part of the intrinsic layer its thickness can be reduced until the radius of the Ag nanosphere. In our final structure the nanospheres are located asymmetrically in the intrinsic layer -- lying on top of the n-doped layer of a-Si. This significant modification of the geometry does not change the power absorption in the intrinsic a-Si. Over the whole frequency range it keeps the same as that shown in Fig. \ref{fig5}(c), green line.

To conclude: in this paper we have studied a collective resonance of a regular array of Ag nanospheres embedded into an optically thin layer of a-Si. 
It was previously shown that this resonance allows this thin metamaterial (metasurface) to  operate as an efficient absorber. In this paper we have shown (via reliable numerical simulations) 
that the very strong absorption holds in a layer of PV material, the regime holds in a practical structure operating as a thin-film PV cell and survives in presence of the metal substrate. 
The most important feature of this light-trapping regime is very low level of parasitic losses in the metal which makes this regime advantageous compared to the anti-reflection regime.
We have found a suitable PV application for our structure. 

Finally, it is worth to note that the practical implementation of our structure can be obtained via the self-assembly of core-shell particles. 
The silver core in any case requires an encapsulation by a solid shell because Ag is chemically active. For dense packaging of the monolayer of such particles the distance between the adjacent spheres (in our notations $p-2r_{\rm Ag}$) equals to the double shell thickness.

\end{document}